\def\Ho{\hat H}
\def\ao{\hat a}

\def\aco{\hat a_c}
\def\r{\rho}\def\ro{\hat \r}
\def\Oo{\hat{\cal O}}
\def\Fo{\hat F}
\def\Jo{\hat J}
\def\jo{\hat j}
\def\l{\lambda}

\documentstyle[12pt]{article}
\begin{document}
\begin{titlepage}
\title{Exact semiclassical wave equation for stochastic quantum optics
\thanks{\it submitted to: Stochastic Quantum Optics, ed.: H. Carmichael
(special issue of Quantum and Semiclassical Optics)}}
\author{Lajos Di\'osi\thanks{E-mail: diosi@rmki.kfki.hu}\\
KFKI Research Institute for Particle and Nuclear Physics\\
H-1525 Budapest 114, POB 49, Hungary\\\\
{\it e-Archive ref.: quant-ph/9507010\hfill}}
\date{Sept 27, 1995}
\maketitle
\begin{abstract}
Semiclassical (stochastic) wave equations are proposed for the coupled
dynamics of atomic quantum states and semiclassical radiation
field. All relevant predictions of standard unitary quantum dynamics
are exactly reproducible in the framework of stochastic wave
equation model. We stress in such a way that the concept of stochastic
wave equations is not to be restricted to the widely used Markovian
approximation.
\end{abstract}
\end{titlepage}
\baselineskip=21pt

{\it 1) Introduction.}
Since many years it has been understood  that one can not
reproduce {\it all}\/ statistical predictions of quantum theory
from pure classical statistics \cite{Bel}. One can do it for
{\it certain}\/ predictions after all.
Reduced dynamics of quantum subsystems (i.e. effective dynamics
of open quantum systems)
can always be interpreted as hybrids of quantum and classical
mechanisms. The corresponding {\it semiclassical}\/ dynamics
(which may on equal footing be called {\it semiquantized}) is
described by {\it stochastic wave equations}\/ (SWEs).
Since ten years they have been attracting
permanent interest in research of foundations \cite{GisDioGhiHalPer}
and, since the 90's \cite{DalDumCar}, in quantum optics, too.

Consider a system of atoms interacting with radiation field.
Basically, all attainable results are approximations: SWEs have been
derived in weak coupling Markovian limit \cite{BelWisGoe}.
In the present work, however,
we investigate the exact unitary dynamics of the atom+radiation system
and we ask the following question.
Can we describe the radiation field in terms of
classical (stochastic) variables instead of quantum ones in such
a manner that all relevant predictions remain {\it exactly}\/ identical
with those of the unitary quantum theory? The answer will be affirmative
in almost all respects. Forthcoming results were anticipated once
\cite{Dio90}
by detailed relativistic calculations. Here we intend to give a short
account on exact SWEs. Readers interested in practical rather than
conceptual aspects may, after reading Section 2, undertand SWEs directly
from Section 5.

{\it 2) The quantum model.}
For simplicity's sake, we consider single
mode (cavity) quantum-electrodynamics.
The Hamiltonian for the composite system of the atom plus radiation field
has the following general form
\begin{equation}
\Ho=\Ho^{ato} +\omega \ao^\dagger\ao+\Jo\ao^\dagger+\Jo^\dagger\ao
\end{equation}
where $\Ho^{ato}$ is the atomic Hamiltonian, $\ao$ is the absorption
operator for the single radiation mode of frequency $\omega$.
In the interaction terms, $\Jo$ is atomic operator proportional
to charges and currents. As usual, we assume that at time $t=0$ the
quantum state of the system is a product of the atomic and cavity
radiation states, respectively:
\begin{equation}
\Psi_0=\psi_0^{ato}\psi_0^{rad}.
\end{equation}
In interaction picture, the quantum state $\Psi$ evolves unitarily:
\begin{equation}
{d\Psi_t\over dt}=-i\Ho_I(t) \Psi_t,
\end{equation}
with interaction Hamiltonian
\begin{equation}
\Ho_I(t)=\jo(t)\ao^\dagger+H.C.
\end{equation}
where $\jo(t)=e^{i\omega t}\Jo(t)$ is the "rotated" version of the
interaction picture current $\Jo(t)$. Though the present work
concerns pure state equations it will nonetheless be convenient
to define a density operator $\ro\equiv\Psi\Psi^\dagger$, too.

{\it 3) Stochastic field variables, semiclassical states and observables.}
The basic field variables $\ao,\ao^\dagger$
do not commute: $[\ao,\ao^\dagger]=1$. We shall, nevertheless,
establish a certain natural correspondence between them and
their classical counterparts $a,a^\ast$.
Let us introduce a special notation for symmetric products:
\begin{equation}
\aco\Oo\equiv
{1\over2}\left(\ao\Oo+\Oo\ao\right)=\{\ao,\Oo\}
\end{equation}
where $\Oo$ is arbitrary operator and $\aco$ is called "superoperator"
(c.f. \cite{Dio90} and references therein). The superoperators
$\aco,\aco^\dagger$ commute with each other \cite{KleSha}:
\begin{equation}
[\aco ,\aco^\dagger]\Oo\equiv\{\ao,\{\ao^\dagger,\Oo\}\}
				 -\{\ao^\dagger,\{\ao,\Oo\}\}=0,
\end{equation}
hence we shall make the natural correspondence, mentioned above,
between the classical complex variable $a$ (or $a^\ast$)
and the superoperator $\aco$ (or $\aco^\dagger$).

We describe the radiation field in terms of the classical
stochastic variables $a,a^\ast$ coupled to the quantized
atomic ones. The state of the quantum system will then be represented
by the {\it semiclassical}\/ "density":
\begin{equation}
\ro(a,a^\ast)
=tr_{rad}\left( \delta(a-\aco)\delta(a^\ast-\aco^\dagger) \ro \right),
\end{equation}
which is expected to be density operator for
the atom and phase space distribution
for the field simultaneously. It should satisfy the positivity condition
$\ro(a,a^\ast)\geq0$ for all complex values of $a,a^\ast$. Then
$tr_{ato}\ro(a,a^\ast)$
is the reduced phase space distribution $\r^{rad}(a,a^\ast)$
of the radiation field mode
and, alternatively, $\int\ro(a,a^\ast)da^\ast da$ is
the reduced density operator $\ro^{ato}$ of the atom.
Similarly to Eq.~(7), we define semiclassical counterpart of
Hermitian observable $\Fo$\/ by:
\begin{equation}
\Fo(a,a^\ast)
=tr_{rad}\left( \delta(a-\aco)\delta(a^\ast-\aco^\dagger) \Fo \right).
\end{equation}
We do however not consider all Hermitian observables but those where
the correspondence is invertible:
\begin{equation}
tr\left(\Fo(\aco,\aco^\dagger)\Oo\right)=tr\left(\Fo\Oo\right)
\end{equation}
for all $\Oo$.
This includes a pretty large class of Hermitian operators.

The equivalence of the semiclassical picture with the full quantized
one relies upon the following equation:
\begin{equation}
tr\left(\Fo(\aco,\aco^\dagger)\ro\right)
= tr_{ato}\int \Fo(a,a^\ast) \ro(a,a^\ast)  da^\ast da
\end{equation}
which follows from Eqs.~(7-9) and from the commutativity of
$\aco$ and $\aco^\dagger$.

{\it 4) The positivity issue and its solution by coarse-graining.}
Let us write the phase space distribution of the radiation field,
defined by tracing Eq.~(7) over the atomic states as well, in
Fourier-representation:
\begin{equation}
\r^{rad}(a,a^\ast)={1\over\pi^2}\int\exp(-\l a^\ast + \l^\ast a)
		tr\left(\exp(-\l^\ast\aco + \l\aco^\dagger)\ro\right)
		d\l^\ast d\l.
\end{equation}
By substituting the identity
\begin{equation}
\exp(-\l^\ast\aco + \l\aco^\dagger)\ro=
\exp(-\l^\ast\ao/2)\exp(\l\ao^\dagger/2)\ro
\exp(\l\ao^\dagger/2)\exp(-\l^\ast\ao/2)
\end{equation}
which follows from the rules (5,6), one can apply the identity
$$\exp(\hat A)\exp(\hat B)=\exp([\hat A,\hat B]/2)\exp(\hat A+\hat B)$$
to obtain
\begin{equation}
\r^{rad}(a,a^\ast)={1\over\pi^2}\int\exp(-\l a^\ast + \l^\ast a)
		tr\left(\exp(-\l^\ast\ao + \l\ao^\dagger)\ro\right)
		d\l^\ast d\l.
\end{equation}
One recognizes that $\r^{rad}(a,a^\ast)$ is identical to the well-known
Wigner-function introduced long ago \cite{Wig}. However, the
Wigner-function can {\it not} be interpreted as phase space density
since it may take negative values as well. From Husimi's work we know
that a minimum coarse-graining will cure the problem \cite{Hus}.
Consequently, we
apply a minimum coarse-graining to the "sharp" distribution
$\ro(a,a^\ast)$ defined by Eq.~(7).
Let $a_0,a_0^\ast$ be noisy complex field variables of vacuum
distribution $(2/\pi)\exp(-2\vert a_0\vert^2)$.
Coarse-grained semiclassical state thus reads:
\begin{equation}
\overline{\ro}(a,a^\ast) =\int
{\exp(-2\vert a_0\vert^2)\over\pi/2}\ro(a+a_0,a^\ast+a_0^\ast)da_0^\ast da_0.
\end{equation}
The corresponding coarse-grained phase space density
\hbox{$\overline{\r}^{rad}(a,a^\ast)=tr\overline{\ro}(a,a^\ast)$}
is identical to the so-called $Q$-function of the radiation field
(see, e.g., in \cite{WalMil}). As is well-known, the Q-function is
$1/\pi$-times the diagonal element of the density operator between
coherent states. Accordingly, the
corresponding relation holds true for $\overline{\ro}(a,a^\ast)$ itself:
\begin{equation}
\overline{\ro}(a,a^\ast)
	={1\over\pi}\langle a,a^\ast\vert \ro \vert a,a^\ast \rangle
\end{equation}
where $\vert a,a^\ast \rangle$ is coherent state of the radiation field
mode. [Two remarks on our notations are in order. First, we indicate the
explicit dependence of coherent states on {\it both}\/ $a$ and $a^\ast$.
Second, the quadratic form above has to be understood on the
factor Hilbert-space of the cavity while the resulting expression is
still an $(a,a^\ast)$-dependent density operator in the atom's
Hilbert-space.] The form (15) guarantees the positivity of the
coarse-grained semiclassical density.

One also introduces coarse-grained semiclassical observables:
\begin{equation}
\overline{\Fo}(a,a^\ast)
=\int{exp(-2\vert a_0\vert^2)\over\pi/2}
\Fo(a+a_0,a^\ast+a_0^\ast)da_0^\ast da_0.
\end{equation}
Then a coarse-grained version of the "sharp" equivalence equation (10)
follows:
\begin{equation}
tr\left(\overline{\Fo}(\aco,\aco^\dagger)\ro \right) =
tr_{ato}\int \Fo(a,a^\ast)\overline{\ro}(a,a^\ast)  da^\ast da,
\end{equation}
i.e. the statistics of coarse-grained quantum observables are still
exactly reproducible from the coarse-grained semiclassical state
$\overline{\ro}(a,a^\ast)$.

One would worry because coarse-grained distributions are believed to
loose information. In typical quantum optics applications things prove
not so bad as long as one considers expectation values of polinoms of
the absorption and emission operators. Then the additional noise is
easy to subtract. Let us assume that we are interested in the
expectation value of the photon number operator $\ao^\dagger\ao$ in the
current quantum state $\ro$ and we want to calculate it from
$\overline{\ro}(a,a^\ast)$.
It is plausible to choose $\Fo(a,a^\ast)=\vert a\vert^2$.
Then, via Eq.~(16), let us calculate the
corresponding coarse-grained observable $\overline{\Fo}(\aco,\aco^\dagger)$.
We obtain $\overline{\aco^\dagger\aco}=\aco^\dagger\aco+{1\over2}$.
The equivalence condition (17) thus leads to:
\begin{equation}
tr\left((\aco^\dagger\aco+{1\over2})\ro\right)
=\int \vert a\vert^2\overline{\r}^{rad}(a,a^\ast)da^\ast da.
\end{equation}
Due to the identity
$tr\left((\aco^\dagger\aco+{1\over2})\ro\right)=tr(\ao^\dagger\ao\ro)+1$,
Eq.~(18) yields:
\begin{equation}
tr(\ao^\dagger\ao\ro)
=\int \vert a\vert^2\overline{\r}^{rad}(a,a^\ast)da^\ast da -1.
\end{equation}
This is the simplest example to illustrate how exact quantum expectation
values are reproduced from the coarse-grained semiclassical
state $\overline{\ro}(a,a^\ast)$.

{\it 5) Exact stochastic wave equation.}
Remind that the composite system of the atom+radiation was originally
assumed to be in a pure quantum state $\Psi$.
Let us introduce a coherent state representation of $\Psi$:
\begin{equation}
\psi(a,a^\ast)={1\over\sqrt{\pi}}\langle a,a^\ast \vert \Psi \rangle.
\end{equation}
The scalar product on the RHS is to be taken on the factor Hilbert-space
of the cavity. [The author has failed to find a more suitable compact
notation, see also the second remark after the Eq.~(15).]
Strictly speaking, $\psi(a,a^\ast)$ is wave function of the field mode
coordinates $a,a^\ast$ while, on the other hand, it is state vector
(of arbitrary representation) in the atomic Hilbert-space.
It follows from Eq.~(15) that the wave function (20) just yields the
coarse-grained semiclassical density in the form
\begin{equation}
\overline{\ro}(a,a^\ast)=\psi(a,a^\ast)\psi^\dagger(a,a^\ast).
\end{equation}
So, it is natural to call $\psi(a,a^\ast)$ semiclassical wave function.
Let us obtain its equation of motion.

In coherent state representation the following operator correspondences
should be used (c.f. \cite{Gar}):
\begin{equation}
\ao^\dagger\Psi\leftrightarrow a^\ast\psi(a,a^\ast),~~~~~
\ao\Psi\leftrightarrow
\left({a\over2}+{\partial\over\partial a^\ast}\right)\psi(a,a^\ast).
\end{equation}
Hence the Hamiltonian equation of motion (3) leads to the following
equation of motion for the semiclassical wave function:
\begin{equation}
{d\psi_t(a,a^\ast)\over dt}=
-i\left(a^\ast\jo(t) + {a\over2}\jo^\dagger(t)\right)\psi_t(a,a^\ast)
-i\jo^\dagger(t){\partial\psi_t(a,a^\ast)\over\partial a^\ast}.
\end{equation}

This wave equation, combining unitary and stochastic dynamics, is our
central result. By means of this equation, exact predictions
of the unitary quantum theory can be otained \cite{foo}.

Let us summarize the SWE method. Assume the atomic state at $t=0$
is $\psi^{ato}_0$ while the cavity's state is the
vacuum state $\vert 0,0\rangle$. Then, according to Eq.~(2), the
semiclassical (stochastic) wave function (20) initially takes the form
\begin{equation}
\psi_0(a,a^\ast)=\psi_0^{ato}{\exp(-\vert a\vert^2/2)\over\sqrt{\pi}}.
\end{equation}
Let us switch on the interaction. Then the conditional atomic wave
function $\psi_t(a,a^\ast)$ evolves in function of the state
$(a,a^\ast)$ of the  classical field, according to the Eq.~(23) while
the (coarse-grained) probability distribution
of the classical field variables $a,a^\ast$ is given by
\begin{equation}
\r_t(a,a^\ast)=\Vert \psi_t(a,a^\ast)\Vert^2.
\end{equation}
The expectation value of an arbitrary semiclassical observable
$\Fo(a,a^\ast)$ will exactly reproduce the quantum expectation value
of the corresponding coarse-grained Hermitian observable, i.e.
\begin{equation}
tr\left(\overline{\Fo}(\aco,\aco^\dagger)\ro_t\right)
=\int \psi_t^\dagger(a,a^\ast)\Fo(a,a^\ast)\psi_t(a,a^\ast) da^\ast da.
\end{equation}
where $\overline{\Fo}(a,a^\ast)$ is derived from $\Fo(a,a^\ast)$ by
the minimum coarse-graining (16).

The existence of exact SWE was pointed out in Ref.~\cite{Dio90}
in context of the relativistic quantum-electrodynamics.
That time, however, the stochastic field was not identified as
semiclassical radiation field. A generalization of our SWE (23) for
infinite number of field modes being initially in thermal equilibrium
states at nonzero temperature seems straightforward.

{\it 6) Outlook.}
The main result of the present work is a suggestion that
semiclassical SWEs could {\it exactly}\/ reproduce the quantum physics of
both atomic {\it and}\/ radiation degrees of freedom. In our opinion, the
very existence of exact stochastic models is a novelty of fundamental
interest. Conceptual elements of equivalence with unitary quantum theory
are symmetrized products of non-commuting operators and a minimum
coarse-garining of states and observables. Our SWE establishes the
possibility of making efficient Monte-Carlo simulations in the
non-Markovian regime since using classical variables for the radiation
field represents a fair simplification. Of course, the re-derivation
of the extensively used Markovian SWEs is possible and desirable.

\bigskip
This work was supported by the grants OTKA No. 1822/1991 and T016047.


\begin{thebibliography}{99}
\bibitem{foo} If we introduce the unnormalized wave function
$\varphi(a^\ast)$ defined by
$\psi(a,a^\ast)=\varphi(a^\ast)\exp(-\vert a\vert^2/2)/\sqrt{\pi}$
then the SWE (23) takes a much simpler form
$${d\varphi_t(a^\ast)\over dt}=
	-ia^\ast\jo(t)\varphi_t(a^\ast)
	-i\jo^\dagger(t){\partial\varphi_t(a^\ast)\over\partial a^\ast}$$
with the initial condition $\varphi_0(a^\ast)=\psi_0^{ato}$ instead of (24).
\end{thebibliography}
\end{document}